\documentclass[letterpaper,12pt]{article}
\usepackage{amssymb}

\newcommand{\leng}[1]{|{#1}|}
\catcode`\@=11
\def\newexample#1{\@ifnextchar[{\@oexm{#1}}{\@nexm{#1}}}

\def\@nexm#1#2{%
\@ifnextchar[{\@xnexm{#1}{#2}}{\@ynexm{#1}{#2}}}

\def\@xnexm#1#2[#3]{\expandafter\@ifdefinable\csname #1\endcsname
{\@definecounter{#1}\@addtoreset{#1}{#3}%
\expandafter\xdef\csname the#1\endcsname{\expandafter\noexpand
  \csname the#3\endcsname \@exmcountersep \@exmcounter{#1}}%
\global\@namedef{#1}{\@exm{#1}{#2}}\global\@namedef{end#1}{\@endexample}}}

\def\@ynexm#1#2{\expandafter\@ifdefinable\csname #1\endcsname
{\@definecounter{#1}%
\expandafter\xdef\csname the#1\endcsname{\@exmcounter{#1}}%
\global\@namedef{#1}{\@exm{#1}{#2}}\global\@namedef{end#1}{\@endexample}}}

\def\@oexm#1[#2]#3{\expandafter\@ifdefinable\csname #1\endcsname
  {\global\@namedef{the#1}{\@nameuse{the#2}}%
\global\@namedef{#1}{\@exm{#2}{#3}}%
\global\@namedef{end#1}{\@endexample}}}

\def\@exm#1#2{\refstepcounter
    {#1}\@ifnextchar[{\@yexm{#1}{#2}}{\@xexm{#1}{#2}}}

\def\@xexm#1#2{\@beginexample{#2}{\csname the#1\endcsname}\ignorespaces}
\def\@yexm#1#2[#3]{\@opargbeginexample{#2}{\csname
       the#1\endcsname}{#3}\ignorespaces}

\def\@exmcounter#1{\noexpand\arabic{#1}}
\def\@exmcountersep{.}
\def\@beginexample#1#2{\trivlist \item[\hskip 
\labelsep{\bf #1\ #2:}]}
\def\@opargbeginexample#1#2#3{\trivlist
      \item[\hskip \labelsep{\bf #1\ #2\ }#3{\bf :}]}
\def\@endexample{\endtrivlist}

\catcode`\@=12

\newtheorem{lemma}{{\bf Lemma}}[section]
\newtheorem{thm}{{\bf Theorem}}[section]
\newtheorem{df}{{\bf Definition}}[section]
\newtheorem{cor}{{\bf Corollary}}[section]

\newtheorem{rem}{{\bf Remark}}[section]

\newtheorem{claim}{{\bf Claim}}[section]

\newcommand{\BQED}{\hfill \hbox{\rule{8pt}{8pt}}}

\newcommand{\ceil}[1]{\lceil{#1}\rceil}
\newcommand{\bm}[1]{\mbox{\boldmath{$#1$}}}
\newcommand{\pair}[1]{\langle{#1}\rangle}
\newcommand{\mod}[1]{\hspace{3pt}({\rm mod}\,\,#1)}

\newenvironment{namelist}[1]{%
\begin{list}{}
  { 
	\settowidth{\labelwidth}{#1}
	\setlength{\leftmargin}{1.1\labelwidth}}
  \setlength{\itemsep}{0cm}
}{%
\end{list}}

\catcode`\@=11
\renewcommand{\@biblabel}[1]{\hspace*{\fill}[#1]}
\catcode`\@=12
\begin{document}
\begin{center}
{\Large {\bf New Constructions for Query-Efficient}}\medskip\\
{\Large {\bf  Locally Decodable Codes of Subexponential Length}}\bigskip\\
\begin{tabular}{ccc}
{\sc Toshiya Itoh} & & {\sc Yasuhiro Suzuki}\\
{\sf titoh@dac.gsic.titech.ac.jp} & & 
{\sf suzuki@dac.gsic.titech.ac.jp}\smallskip\\
{\sf Global Scientific and Computing Center} & & {\sf Department of Computer Science}\\
{\sf Tokyo Institute of Technology} & & {\sf Tokyo Institute of Technology}\\
{\sf Meguro-ku, Tokyo 152-8550, Japan} & & 
{\sf Meguro-ku, Tokyo 152-8550, Japan}
\end{tabular}
\end{center}\medskip\medskip
{\bf Abstract:} A $(k,\delta,\varepsilon)$-locally decodable code 
$C:{\bf F}_{q}^{n} \rightarrow {\bf F}_{q}^{N}$ is an 
error-correcting code that encodes each message 
$\vec{x}=(x_{1},x_{2},\ldots,x_{n}) \in {\bf F}_{q}^{n}$ to a codeword 
$C(\vec{x}) \in {\bf F}_{q}^{N}$ and has the following property:~For 
any $\vec{y} \in {\bf F}_{q}^{N}$ such that 
$d(\vec{y},C(\vec{x})) \leq \delta N$ and 
each $1 \leq i \leq n$, the symbol $x_{i}$ of $\vec{x}$ can be 
recovered~with probability at least $1-\varepsilon$ 
by a randomized decoding algorithm looking only at $k$ coordinates of $\vec{y}$.~The 
efficiency of a $(k,\delta,\varepsilon)$-locally decodable code 
$C:{\bf F}_{q}^{n} \rightarrow {\bf F}_{q}^{N}$ is measured by 
the code length~$N$~and~the number $k$ of queries. 
For any $k$-query 
locally decodable code $C:{\bf F}_{q}^{n} \rightarrow {\bf F}_{q}^{N}$, 
the code length~$N$~is~con\-jectured to be exponential of $n$, i.e., 
$N=\exp(n^{\Omega(1)})$, however, this was disproved.~Yekhanin~[In~Proc. 
of STOC, 2007] showed that there exists a 3-query locally decodable code 
$C:{\bf F}_{2}^{n} \rightarrow {\bf F}_{2}^{N}$~such~that 
$N=\exp(n^{(1/\log \log n)})$ 
assuming that the number of Mersenne 
primes is infinite. For a 3-query locally decodable code 
$C:{\bf F}_{q}^{n} \rightarrow {\bf F}_{q}^{N}$, 
Efremenko [ECCC Report No.69,~2008]~reduced~the~code~length~fur\-ther to 
$N=\exp(n^{O((\log \log n/ \log n)^{1/2})})$, and also showed that
for any integer $r>1$,~there~exists~a~$k\mbox{-}$que\-ry locally decodable code 
$C:{\bf F}_{q}^{n} \rightarrow {\bf F}_{q}^{N}$ 
such that $k \leq 2^{r}$ and 
$N=\exp(n^{O((\log \log n/ \log n)^{1-1/r})})$.~In this paper, we 
present a query-efficient locally decodable code by introducing 
a technique of ``composition of locally decodable codes,'' and 
show that for any integer $r>1$, 
there exists a $k$-query locally~decod\-able code 
$C:{\bf F}_{q}^{n} \rightarrow {\bf F}_{q}^{N}$ 
such that $k \leq 3 \cdot 2^{r-2}$ and 
$N=\exp(n^{O((\log \log n/ \log n)^{1-1/r})})$.\medskip

\section{Introduction} \label{intro}
%
Conventional error-correcting codes 
$C: {\bf F}_{q}^{n} \rightarrow {\bf F}_{q}^{N}$ allow one to encode any 
$\vec{x} = (x_{1},x_{2},\ldots,x_{n})\in {\bf F}_{q}^{n}$ to 
$C(\vec{x}) \in {\bf F}_{q}^{N}$ and 
have the following property: For any $\vec{y} \in {\bf F}_{q}^{N}$ such 
that $d(\vec{y},C(\vec{x})) \leq \delta N$,~the~orig\-inal message 
$\vec{x}$ can be recovered by looking at entire 
coordinates of $\vec{y}$. If one is interested in recovering a single 
symbol $x_{i}$ of $\vec{x}$, more efficient schemes are possible. Such 
schemes are~known~as~{\it locally~decoda\-ble codes\/} 
$C:{\bf F}_{q}^{n} \rightarrow {\bf F}_{q}^{N}$ 
that allow recovery of 
any single symbol $x_{i}$ of $\vec{x} \in {\bf F}_{q}^{n}$ 
by looking only~at~$k$~ran\-domly 
chosen coordinates of $\vec{y} \in {\bf F}_{q}^{N}$ 
such that $d(\vec{y},C(\vec{x}))\leq \delta N$. 
Informally, a $(k,\delta,\varepsilon)$-locally~decod\-able code  
$C:{\bf F}_{q}^{n} \rightarrow {\bf F}_{q}^{N}$ 
is an 
error-correcting code that encodes each message 
$\vec{x}=(x_{1},x_{2},\ldots,x_{n}) \in {\bf F}_{q}^{n}$ to a codeword 
$C(\vec{x}) \in {\bf F}_{q}^{N}$ and has the following property: For 
any $\vec{y} \in {\bf F}_{q}^{N}$ such that 
$d(\vec{y},C(\vec{x})) \leq \delta N$ and 
each $1 \leq i \leq n$, the symbol $x_{i}$ of $\vec{x}$ can be 
recovered with probability at least $1-\varepsilon$ 
by a randomized decoding algorithm looking only at $k$ coordinates of $\vec{y}$. 
%
\subsection{Known Results} \label{known}
%
From theoretical and practical point of view, 
we are interested in designing a $(k,\delta,\varepsilon)$-locally~decoda\-ble 
code $C: {\bf F}_{q}^{n} \rightarrow {\bf F}_{q}^{N}$ 
as shorter $N$ as possible and as smaller $k$ as possible. 
The~notion~of~locally~de\-codable codes was considered in several contexts 
\cite{BFLS,S,PS}, and Katz and Trevisan \cite{KT}~were~the~first to 
provide a formal definition of locally decodable codes 
and prove lower bounds for the~code~length. 
Gasarch 
\cite{Ga} and Goldreich 
\cite{Go} conjectured that for a $k$-query 
locally decodable~code~$C: {\bf F}_{q}^{n} \rightarrow {\bf F}_{q}^{N}$ 
with $k>1$, the code length $N$ is unavoidable to be the 
exponential~of~$n$,~i.e.,~$N=\exp(n^{\Omega(1)})$.~In~Table \ref{tab-known-length}, 
we summarize the known results on the code length for 
$k$-query locally decodable codes. 

\begin{table}[htb]
\begin{center}
\caption{Known Results on the Code Length}\medskip \label{tab-known-length}
\def\arraystretch{1.7}
\begin{tabular}{|c||cl|cl|} \hline
        & \multicolumn{2}{c|}{Upper Bound}
& \multicolumn{2}{c|}{Lower Bound}\\ \hline\hline
\makebox[2.0cm]{2-Query} 
& \makebox[4.25cm]{$\exp\left(O(n)\right)$} & \cite{KW} 
& \makebox[4.25cm]{$\exp\left(\Omega(n)\right)$} & \cite{KW}\\ \hline
3-Query & $\exp\left(n^{1/2}\right)$ & \cite{BIK} 
& $\tilde{\Omega}\left(n^{2}\right)$ & \cite{KW,W}\\ \hline
$k$-Query & $\exp\left(n^{O(\log \log k)/k \log k}\right)$ & \cite{BIKR} 
& $\tilde{\Omega}\left(n^{1+1(\ceil{k/2}-1)}\right)$ & \cite{KW,W}\\ \hline
\end{tabular}
\end{center}
\end{table}

\noindent Yekhanin \cite{Y1,Y2} improved the upper bound 
for the code length of 3-query locally 
decodable~codes~to $N=\exp(n^{1/32582657})$  
and disproved the conjecture \cite{Ga,Go} 
on the code length of 3-query locally~decod\-able codes, i.e., 
if there exist infinitely many Mersenne primes, then 
$N=\exp(n^{O(1/\log \log n)})$~for~infi\-nitely many $n$'s. 
Very recently, Efremenko \cite[Theorem 3.8]{E} 
improved much further the upper bound for the code 
length of 3-query locally decodable codes to
\[
N=\exp\left(\exp\left(O\left(\sqrt{\log n \cdot \log \log n}\right)\right)\right)
=\exp\left(n^{O(\left(\log \log n/\log n\right)^{1/2})}\right),
\]
by introducing the notions of $S$-matching vectors \cite[Definition 3.1]{E} 
and $S$-decoding polynomials \cite[Definition 3.4]{E} --- 
this reduces the code length of 3-query locally decodable codes 
and removes~the~unproven assumption that 
infinitely many Mersenne primes exist. 
For any $k>2$, Efremenko \cite[Theorem 3.6]{E} 
also disproved the conjecture \cite{Ga,Go} on the code length of $k$-query 
locally decodable codes, and showed that 
for any $r>1$, there exists a $k$-query locally decodable code such that 
$k \leq 2^{r}$ and 
\[
N=\exp\left(\exp\left(O\left(\sqrt[r]{ \log n \cdot 
\left(\log \log n\right)^{r-1}}\right)\right)\right)
=\exp\left(n^{O(\left(\log \log n/\log n\right)^{1-1/r})}\right). 
\]
%
\subsection{Main Result} \label{main}
%
In this paper, we present an improved construction of a $k$-query locally 
decodable code $C:{\bf F}_{q}^{n} \rightarrow {\bf F}_{q}^{N}$, and 
show that for any $r>1$, there exists a 
$k$-query locally decodable code such that $k \leq 3\cdot 2^{r-2}$~and 
\[
N=\exp\left(\exp\left(O\left(\sqrt[r]{ \log n \cdot \left(\log \log n
\right)^{r-1}}\right)\right)\right)
=\exp\left(n^{O(\left(\log \log n/\log n\right)^{1-1/r})}\right).  
\]
Our construction of the $3\cdot 2^{r-2}$-query locally decodable codes 
is partially based on the construction~by Efremenko \cite{E}. To reduce 
the number of queries, we introduce a technique of ``composition of 
locally decodable codes.'' In fact, we show that for 
a $k_{1}$-query locally decodable code and a 
$k_{2}$-query~locally decodable code, there exists a $k_{1}k_{2}$-query 
locally decodable code. 
Applying~our~technique~of~``compo\-sition of 
locally decodable codes'' to the 3-query locally decodable code 
\cite[Theorem 3.8]{E}~and~the~$2^{r-2}$-query locally decodable code 
\cite[Theorem 3.6]{E}, a $3\cdot 2^{r-2}$-query locally decodable code is 
achieved. 
%
\subsection{Application of Locally Decodable Codes} \label{application}
%
Locally decodable codes have many applications in complexity theory 
and cryptography (see, e.g., \cite{T,Ga}). In particular, locally decodable codes 
are closely related to designing efficient private~infor\-mation retrieval. 
Informally, a $k$-server private information retrieval 
is a protocol that consists~of~a user ${\cal U}$ and 
$k$ databases ${\cal DB}_{1},{\cal DB}_{2},\ldots,{\cal DB}_{k}$ 
with identical data $\vec{x}=(x_{1},x_{2},\ldots,x_{n})$, 
where~each~data\-base ${\cal DB}_{j}$ does not communicate to any other 
database ${\cal DB}_{h}$, and allows 
the user ${\cal U}$ to retrieve~$x_{i}$~of~$\vec{x}$ while any of the $k$ 
databases ${\cal DB}_{1},{\cal DB}_{2},\ldots,{\cal DB}_{k}$ 
learns nothing about $i$. Private information~retriev\-al 
was introduced by Chor et al. \cite{CGKS}, 
and the efficiency of a $k$-server private~information~retrieval~is 
measured by its communication complexity $C_{k}(n)$, i.e., 
the total amount of bits exchanged between the user ${\cal U}$ and each of 
the $k$ databases ${\cal D}_{1},{\cal D}_{2},\ldots,{\cal D}_{k}$. 
For further details on $k$-server private~infor\-mation retrieval, 
see, e.g., \cite{A,M,I1,I2,GKST,KW,BIK,BFG,RY,WY}. 

\begin{table}[htb]
\begin{center}
\caption{Known Results on the Communication Complexity}\medskip \label{tab-known-cc}
\def\arraystretch{1.5}
\begin{tabular}{|c||cl|cl|} \hline
        & \multicolumn{2}{c|}{Upper Bound}
& \multicolumn{2}{c|}{Lower Bound}\\ \hline\hline
\makebox[2.0cm]{1-Server} 
& \makebox[3.5cm]{$n+1$} & \cite{CGKS} 
& \makebox[2.25cm]{$n$} & \cite{CGKS}\\ \hline
2-Server & $n^{1/3}$ & \cite{CGKS,IK} 
& $5\log n$ & \cite{WW}\\ \hline
3-Server & $n^{O((\log \log n/\log n)^{1/2})}$ & \cite{E} 
& \multicolumn{2}{c|}{---}\\ \hline
4-Server & $n^{1/7.87}$ & \cite{BIKR} & \multicolumn{2}{c|}{---}\\ \hline
$k$-Server & $n^{O(\log \log k/k\log k)}$ & \cite{BIKR} & 
\multicolumn{2}{c|}{---}\\ \hline
\end{tabular}
\end{center}
\end{table}

In Table \ref{tab-known-cc}, 
we summarize the known results on the communication complexity $C_{k}(n)$ 
for $k$-server private information retrieval. In particular, 
Efremenko \cite[Theorem 3.6]{E} showed~that~a~communica\-tion-efficient 
$k$-server private information retrieval exists for a specific $k>1$, i.e., 
for~any~$r>1$,~there exists 
a $k$-server private information retrieval such that $k \leq 2^{r}$ and 
$C_{k}(n)=n^{O((\log \log n/\log n)^{(r-1)/r})}$. 
%
\section{Preliminaries} \label{preliminary}
%
\subsection{Locally Decodable Codes} \label{ldc}
%
We use ${\bf F}_{q}$ to denote a finite field of $q$ elements and 
$d(\vec{x},\vec{y})$ to denote the Hamming distance~of~vectors 
$\vec{x} = (x_{1},x_{2},\ldots, x_{n}) \in {\bf F}_{q}^{n}$ and 
$\vec{y} = (y_{1},y_{2},\ldots, y_{n}) \in {\bf F}_{q}^{n}$, i.e., 
the number of indices such that $x_{i}\neq y_{i}$. 
For any integer $a<b$, we use $[a,b]$ to denote the set $\{a,a+1,\ldots,b\}$. 
For any integer~$m>1$,~let~${\bf Z}_{m}=\{0,1,\ldots,m-1\}$ 
and ${\bf Z}_{m}^{*} =\{z \in {\bf Z}_{m}: \gcd(z,m)=1\}$. 
\begin{df}[\cite{KT}] \label{def-LDC}
We say that $C:{\bf F}_{q}^{n} \rightarrow {\bf F}_{q}^{N}$ is a 
{\sf $(k,\delta,\varepsilon)$-locally decodable code} 
if~for~each~$i \in [1,n]$, 
there exists a randomized decoding algorithm 
$D_{i}: {\bf F}_{q}^{N} \rightarrow {\bf F}_{q}$ such that {\rm (1)} for 
any message~$\vec{x} = (x_{1},x_{2},\ldots,x_{n}) \in {\bf F}_{q}^{n}$ and any 
$\vec{y} \in {\bf F}_{q}^{N}$ such that $d(C(\vec{x}),\vec{y}) \leq \delta N$,  
$\Pr[D_{i}(\vec{y})=x_{i}] \geq 1-\epsilon;$ 
{\rm (2)} the algorithm $D_{i}$ makes at most $k$ queries to $\vec{y}$. 
\end{df}

We say that a $(k,\delta,\epsilon)$-locally decodable code $C$ 
is {\it linear\/} if $C$ is linear 
over ${\bf F}_{q}$ and is {\it nonadaptive\/} if for each $i \in [1,n]$, 
the decoding algorithm $D_{i}$ makes all its queries simultaneously. 
In~this~paper,~we deal with only linear and nonadaptive 
$(k,\delta,\epsilon)$-locally decodable codes. 
\begin{df}[\cite{T}] \label{def-smooth}
We say that $C:{\bf F}_{q}^{n} \rightarrow {\bf F}_{q}^{N}$ has a 
{\sf perfectly smooth decoder} ${\cal D}= \{D_{i}\}_{i \in [1.n]}$ if for 
each $\vec{x} \in {\sf F}_{q}^{n}$ 
and each $i \in [1,n]$, 
$\Pr[D_{i}(C(\vec{x}))=x_{i}]=1$, and 
each query made~by~the~randomized~de\-coding algorithm $D_{i}$ 
is uniformly distributed over $[1,N]$. 
\end{df}
Trevisan \cite{T} observed that for a code 
$C:{\bf F}_{q}^{n} \rightarrow {\bf F}_{q}^{N}$, if $C$ has a perfectly 
smooth decoder~and~makes at most $k$ queries, 
then $C$ is a $(k,\delta,k\delta)$-locally 
decodable code. Thus in the rest of this paper,~we~use $k$-query 
locally decodable codes instead of 
$(k,\delta,\varepsilon)$-locally decodable codes. 
%
\subsection{{\bm S}-Matching Vectors} \label{S-matching}
%
Let $m> 1$ and $h>0$ be integersD
For any 
$\vec{x} = (x_{1},x_{2},\ldots, x_{h})\in {\bf Z}_{m}^{h}$~and 
$\vec{y} = (y_{1},y_{2},\ldots, y_{h}) \in {\bf Z}_{m}^{h}$, we use 
$\pair{\vec{x}, \vec{y}}_{m}$ 
to denote the {\it inner product\/} of $\vec{x}$ and $\vec{y}$ modulo $m$,~i.e., 
\[
\pair{\vec{x},\vec{y}}_{m} \equiv \sum_{j=1}^{h} x_{j}y_{j} \mod{m}.
\]
\begin{df}[\cite{E}] \label{def-S-matching}
Let $S \subseteq {\bf Z}_{m} \setminus \{0\}$ and 
${\cal U}=\{\vec{u}_{1},\vec{u}_{2},\ldots, \vec{u}_{n}\}$ be a family of 
vectors,~where~$\vec{u}_{i} \in {\bf Z}_{m}^{h}$ for each $i \in [1,n]$. 
We say that a family ${\cal U}=\{\vec{u}_{1},\vec{u}_{2},\ldots, \vec{u}_{n}\}$ 
of vectors is {\sf $S$-matching} if 
{\rm (1)}~for~each $i \in [1,n]$, $\pair{\vec{u}_{i},\vec{u}_{i}}_{m}=0;$ 
{\rm (2)} for each $i,j \in [1,n]$ such that $i \neq j$, 
$\pair{\vec{u}_{i},\vec{u}_{j}}_{m} \in S$. 
%
%
%
%
%
\end{df}
Let $m=p_{1}^{e_{1}}p_{2}^{e_{2}} \cdots p_{r}^{e_{r}}$ be a product of 
$r>1$ distinct primes. Define $S_{m} \subseteq {\bf Z}_{m}\setminus \{0\}$ as 
follows:~For~each $s \in {\bf Z}_{m} \setminus \{0\}$, if either 
$s \equiv 0 \mod{p_{i}^{e_{i}}}$ or $s \equiv 1 \mod{p_{i}^{e_{i}}}$ 
for each $i \in [r]$, then $s \in S_{m}$. We refer~to~$S_{m}$ as the {\it canonical\/} 
set of the integer $m=p_{1}^{e_{1}}p_{2}^{e_{2}} \cdots p_{r}^{e_{r}}$. 

For each integer $t \in [0,2^{r}-1]$, we use 
${\rm bin}(t)=(t_{r-1},t_{r-2},\ldots ,t_{0}) \in \{0,1\}^{r}$ to 
denote the~binary representation of $t$, i.e., 
$t=t_{r-1}\cdot 2^{r-1}+t_{r-2}\cdot 2^{r-2}+\cdots + t_{0}\cdot 2^{0}$, 
and let $s_{t} \in [0,m-1]$ be an integer such that 
$s_{t} \equiv t_{i-1} \mod{p_{i}^{e_{i}}}$ for each $i \in [1,r]$. 
Thus from the definition of $S_{m} \subseteq Z_{m}\setminus \{0\}$,~it~follows 
that $S_{m}=\{s_{1},s_{2}.\ldots, s_{2^{r}-1}\}$, where 
$s_{0}=0$ and $s_{2^{r}-1}=1$. 
%
\begin{lemma}[\mbox{\cite[Theorems 1.2 and 1.3]{G}}] \label{lemma-G}
Let $m=p_{1}^{e_{1}}p_{2}^{e_{2}} \cdots p_{r}^{e_{r}}$ be a product of 
$r>1$ distinct~primes. Then there exists a constant $c=c(m)>0$ such that 
for every integer $h>0$, there exists an explicitly constructible 
uniform set-system ${\cal H}$ over the universe $[1,h]$ that 
satisfies the following$:$
\begin{namelist}{~~~(3)}
\item[{\rm (1)}] $\leng{{\cal H}} \geq 
\exp\left(c\frac{(\log h)^{r}}{(\log \log h)^{r-1}}\right);$
\item[{\rm (2)}] for each $H \in {\cal H}$, $\leng{H}\equiv 0 \mod{m};$
\item[{\rm (3)}] for any $G,H \in {\cal H}$ such that $G \neq H$, 
there exists $i \in [1,2^{r}-1]$ such that 
$\leng{G \cap H} \equiv s_{i} \mod{m}$, where 
$S_{m}=\{s_{1},s_{2},\ldots,s_{2^{r}-1}\}$ is the canonical set of $m$.
\end{namelist}
\end{lemma}
For each $H_{i} \in {\cal H}$, let $\vec{u}_{i} 
=(u_{i1},u_{i2},\ldots, u_{ih})\in \{0,1\}^{h}$ 
be the incidence vector of $H_{i}$,~i.e.,~for~each~$j \in [1,h]$, 
$u_{ij} =1$ iff $j \in H_{i}$. 
By Lemma \ref{lemma-G}, Efremenko \cite{E} showed the following results: 
\begin{lemma}[\mbox{\cite[Corollary 3.3]{E}}] \label{lemma-E}
Let $m=p_{1}^{e_{1}}p_{2}^{e_{2}} \cdots p_{r}^{e_{r}}$ be a product of 
$r>1$ distinct~primes and~$S_{m}$ be the canonical set of $m$. 
Then for any integer $h>0$, there exists a family 
${\cal U}=\{\vec{u}_{1},\vec{u}_{2},\ldots, \vec{u}_{n}\}$~of $S_{m}$-matching 
vectors such that $\vec{u}_{i} \in \{0,1\}^{h} \subseteq 
{\bf Z}_{m}^{h}$ for each $i \in [n]$ and 
$n \geq \exp\left(c\frac{(\log h)^{r}}{(\log \log h)^{r-1}}\right)$. 
\end{lemma}
%
\subsection{\bm{S}-Decoding Polynomials} \label{S-decoding}
%
To construct a $(k,\delta,\epsilon)$-locally decodable codes of 
short length, the following lemma is useful. 
\begin{lemma}[\mbox{\cite[Fact 2.4]{E}}] \label{lemma-finite}
For any odd integer $m>1$, there exist a finite field ${\bf F}_{2^{t}}$ 
with $t \in [1,m-1]$ and an element $\gamma \in {\bf F}_{2^{t}}$ of order $m$, 
i.e., $\gamma^{m}=1$ and $\gamma^{i} \neq 1$ for each $i \in [1,m-1]$.
\end{lemma}
%
Let $m=p_{1}^{e_{1}}p_{2}^{e_{2}} \cdots p_{r}^{e_{r}}$ be a product of $r>1$ 
distinct odd primes and $\gamma \in {\bf F}_{2^{t}}$ be 
an element given~by Lemma \ref{lemma-finite}. 
Efremenko~\cite{E} introduced a notion of $S$-decoding polynomials, which 
plays~a~crucial~role to construct a query-efficient locally decodable code. 
\begin{df}[\mbox{\cite[Definition 3.4]{E}}] \label{def-S-decoding}
For any $S \subseteq {\bf Z}_{m}\setminus \{0\}$, 
we say that $P(x) \in {\bf F}_{2^{t}}[x]$ 
is~an~{\sf $S$-decod\-ing~polynomial} if {\rm (1)} $P(\gamma^{s})=0$ 
for each $s \in S;$ {\rm (2)} $P(\gamma^{0})=P(1)=1$.
\end{df}
Efremenko \cite{E} showed that there exists an $S$-decoding polynomial 
with a few monomials. 
\begin{lemma}[\mbox{\cite[Claim 3.1]{E}}] \label{lemma-S-decoding}
For any odd integer $m=p_{1}^{e_{1}}p_{2}^{e_{2}} \cdots p_{r}^{r_{r}}$ with $r>1$ 
and any $S \subseteq {\bf Z}_{m}\setminus \{0\}$, 
there exists an $S$-decoding polynomial $P(x)$ 
with at most $\leng{S}+1$ monomials. 
\end{lemma}
\begin{rem} \label{remark}
The number of monomials of an $S$-decoding polynomial is closely 
related~to~the~num\-ber of queries of the corresponding 
locally decodable code. In fact, 
the number of monomials~of~an~$S$-decoding polynomial is $k$ iff the number of 
queries of the corresponding locally decodable code~is~$k$. 
\end{rem}

Let $m=p_{1}^{e_{1}}p_{2}^{e_{2}} \cdots p_{r}^{e_{r}}$ be a product of 
$r>1$ distinct odd primes. 
It is immediate that $\leng{S_{m}}=2^{r}-1$ 
from the definition of the canonical set $S_{m}$ of $m$. 
Thus from Lemma \ref{lemma-S-decoding}, 
we have the following~lemma: 
\begin{lemma}[\cite{E}] \label{lemma-2^{r}}
Let $m=p_{1}^{e_{1}}p_{2}^{e_{2}} \cdots p_{r}^{e_{r}}$ be a product of 
$r>1$ distinct odd primes. Then~there~exists an $S_{m}$-decoding polynomial 
$P(x)$ with at most $2^{r}$ monomials. 
\end{lemma}
%
\section{Known Construction for $\bm{k}$-Locally Decodable Codes} 
\label{know-ldc}
%
We describe the construction of $(k,\delta,\varepsilon)$-locally decodable codes
given~by Efremenko \cite{E}. 
%
\subsection{Encoding} \label{encoding-efremenko}
%
Let $m=p_{1}^{e_{1}}p_{2}^{e_{2}} \cdots p_{r}^{e_{r}}$ be a product of 
$r>1$ distinct odd primes, $\gamma \in {\bf F}_{2^{t}}$ be an 
element~determined~by Lemma \ref{lemma-finite}, 
and $P(x) =a_{0}+a_{1}x^{b_{1}}+\cdots + a_{k-1}x^{b_{k-1}}
\in {\bf F}_{2^{t}}[x]$ be an $S_{m}$-decoding polynomial,~where~$S_{m}$ 
is the canonical set of $m$. 
For each $i \in [1,n]$, 
let $\vec{e}_{i} \in {\bf F}_{2^{t}}^{n}$ be the $i$th unit vector and 
$N=m^{h}$, where 
\begin{equation}
h = \exp\left(O\left(\sqrt[r]{(\log n)\cdot  (\log \log n)^{r-1}}\right)\right)
= n^{O((\log \log n/\log n)^{1-1/r})}. \label{eq-h}
\end{equation}
Let ${\cal U}=\{\vec{u}_{1},\vec{u}_{2},\ldots,\vec{u}_{n}\}$ be a family of 
$S_{m}$-matching vectors, where $\vec{u}_{i} \in {\bf Z}_{m}^{h}$ 
for each $i \in [1,n]$.~We define a code 
$C: {\bf F}_{2^{t}}^{n} \rightarrow {\bf F}_{2^{t}}^{N}$ 
as follows: 
For any $\vec{x}=(x_{1},x_{2},\ldots,x_{n}) \in 
{\bf F}_{2^{t}}^{n}$,~let~$C(\vec{x}) = 
x_{1} C(\vec{e}_{1})+ x_{2} C(\vec{e}_{2})+ \cdots + x_{n} C(\vec{e}_{n})$, 
where for each $i \in [1,n]$, 
\begin{equation}
C(\vec{e}_{i}) = 
\left(\gamma^{\pair{\vec{u}_{i},\vec{z}}_{m}} \right)_{\vec{z} \in {\bf Z}_{m}^{h}}. 
\label{eq-encoding}
\end{equation}
%
\subsection{Decoding} \label{decoding-efremenko}
\begin{figure}[htb]
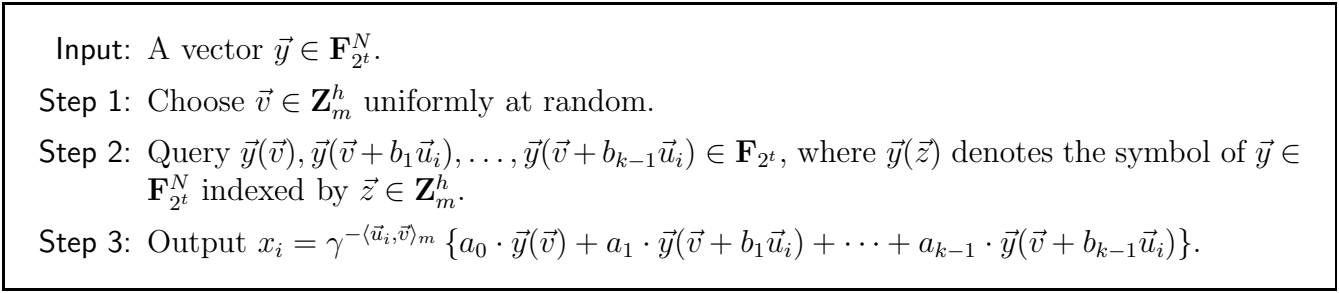

\fbox{
\begin{minipage}{17.35cm}
\begin{minipage}{17.0cm}\smallskip
\begin{namelist}{~~{\sf Step 3:}}
\item[{\sf Input:}] A vector $\vec{y} \in {\bf F}_{2^{t}}^{N}$.
\item[{\sf Step 1:}] Choose $\vec{v} \in {\bf Z}_{m}^{h}$ uniformly at random. 
\item[{\sf Step 2:}] Query $\vec{y}(\vec{v}), 
\vec{y}(\vec{v}+b_{1}\vec{u}_{i}), \ldots, 
\vec{y}(\vec{v}+b_{k-1}\vec{u}_{i}) \in {\bf F}_{2^{t}}$, where $\vec{y}(\vec{z})$ 
denotes the symbol~of~$\vec{y} \in {\bf F}_{2^{t}}^{N}$ indexed by 
$\vec{z} \in {\bf Z}_{m}^{h}$. 
\item[{\sf Step 3:}] Output $x_{i} = \gamma^{-\pair{\vec{u}_{i},\vec{v}}_{m}}
\left\{a_{0} \cdot \vec{y}(\vec{v}) 
+ a_{1} \cdot \vec{y}(\vec{v}+b_{1}\vec{u}_{i}) + 
\cdots + a_{k-1}\cdot \vec{y}(\vec{v}+b_{k-1}\vec{u}_{i})\right\}$. 
\end{namelist}
\end{minipage}\\[0.25cm]
\end{minipage}}
\caption{Decoding Algorithm $D_{i}$} \label{fig-decoding}
\end{figure}

%
For each $i \in [1,n]$, a randomized decoding algorithm 
$D_{i}: {\bf F}_{2^{t}}^{N} \rightarrow {\bf F}_{2^{t}}$ is defined as 
in Figure \ref{fig-decoding}. 

\begin{lemma}[\mbox{\cite[Lemma 3.5]{E}}] \label{lemma-smooth}
The decoding algorithm ${\cal D}=\{D_{i}\}_{i \in [1,n]}$ 
is a perfectly smooth decoder. 
\end{lemma}
To be self-contained, we show the proof of Lemma \ref{lemma-smooth} 
in Appendix \ref{appendix-proof-smooth}. Thus from Lemmas \ref{lemma-E} 
and \ref{lemma-2^{r}}, we have the following result: 
\begin{thm}[\mbox{\cite[Theorem 3.6]{E}}] \label{thm-subexp}
For any integer $n>1$ and any integer $r>1$, 
there exists a $k$-query locally decodable code 
$C: {\bf F}_{2^{t}}^{n} \rightarrow {\bf F}_{2^{t}}^{N}$ such that 
$k \leq 2^{r}$ and 
\[
N=\exp\left(\exp\left(O\left(\sqrt[r]{ \log n \cdot 
\left(\log \log n\right)^{r-1}}\right)\right)\right)
=\exp\left(n^{O(\left(\log \log n/\log n\right)^{1-1/r})}\right). 
\]
\end{thm}
%
\section{Query-Efficient Locally Decodable Codes} \label{new-construction}
%
\subsection{How to Reduce the Number of Queries} \label{how-to}
%
By setting $r=2$ in Theorem \ref{thm-subexp}, it is immediate to see that 
for any integer $n>1$, there exists~a~4-query locally decodable code 
$C:{\bf F}_{2^{t}}^{n} \rightarrow {\bf F}_{2^{t}}^{N}$ such that 
\begin{equation}
N=\exp\left(\exp\left(O\left(\sqrt{ \log n \cdot 
\log \log n}\right)\right)\right)
=\exp\left(n^{O(\left(\log \log n/\log n\right)^{1/2})}\right). 
\label{eq-length-3}
\end{equation}
On the other hand, Efremenko \cite[Example 3.7]{E} found a surprising example: 
Let $m=511=2^{9}-1=7 \cdot 73$ and $S_{511}=\{1,365,147\}$. 
For the integer $m=511$, determine a finite field ${\bf F}_{2^{t}}$ 
and an element $\gamma \in {\bf F}_{2^{t}}$ of order $m=511$ by 
Lemma \ref{lemma-finite}. Indeed, the finite field ${\bf F}_{2^{t}}$ is 
${\bf F}_{2^{9}}= {\bf F}_{2}[\gamma]/(\gamma^{9}+\gamma^{4}+1)$~and 
$\gamma \in {\bf F}_{2^{9}}$ is an element of order 511. For the integer 
$m=511$, there exists an $S_{511}$-decoding~polynomial 
$P(x)=\gamma^{423}\cdot x^{65}+\gamma^{257}\cdot x^{12}+\gamma^{342}$ with 
3 monomials, which implies that for any $n>1$, 
there~exists~a~3-query locally decodable code 
$C:{\bf F}_{2^{9}}^{n} \rightarrow {\bf F}_{2^{9}}^{N}$, where $N$ is given by 
(\ref{eq-length-3}). 

The result above for the integer $m=511$ is special. For an integer 
$m=15=2^{4}-1=3 \cdot 5$, let $S_{15}=\{1,10,6\}$ and by Lemma \ref{lemma-finite}, 
we take the finite field ${\bf F}_{2^{t}}$ to be 
${\bf F}_{2^{4}}={\bf F}_{2}[\gamma]/(\gamma^{4}+\gamma+1)$~and~the 
element $\gamma \in {\bf F}_{2^{4}}$ of order 15. 
By an exhaustive search, we can verify that 
for the integer~$m=15$,~there does not exist an $S_{15}$-decoding polynomial 
with less than 4 monomials.~From~these~observations,~we 
see that it is impossible for every odd integer 
$m=p_{1}^{e_{1}}p_{2}^{e_{2}}$ to have 
an $S_{m}$-decoding~polynomial~with less than 4 monomials. 
Thus for an odd integer $m=p_{1}^{e_{1}}p_{2}^{e_{2}}\cdots p_{r}^{e_{r}}$, 
we need to find structural~properties of $S_{m}$-decoding polynomials 
to reduce the number of queries to less than $2^{r}$. 
%
\subsection{Building Blocks for Query-Efficient Locally Decodable Codes} 
\label{building}
%
In this section, we present a new construction for query-efficient 
locally decodable codes~of~subexpo\-nential length. 
A key idea of our construction is to generate a 
$k_{1}k_{2}$-locally decodable~code~by~compos\-ing a $k_{1}$-locally decodable code 
and a $k_{2}$-locally decodable code. 

Let $m_{1}=p_{1}^{e_{1}}p_{2}^{e_{2}} \cdots p_{r}^{e_{r}}$ be a product 
of $r>1$ distinct odd primes and 
$m_{2}=q_{1}^{c_{1}}q_{2}^{c_{2}} \cdots q_{\ell}^{c_{\ell}}$~be~a~product 
of $\ell>1$ distinct odd primes. Assume that $\gcd(m_{1},m_{2})=1$ 
in the rest of this paper~and~let~$m=m_{1}m_{2}$ be a product of 
$r+\ell>2$ distinct odd primes. From Lemma \ref{lemma-finite}, we know that 
(1)~for~the~odd integer $m_{1}$, 
there exist a finite field ${\bf F}_{2^{t_{1}}}$ with $t_{1} \in [1,m_{1}-1]$ 
and an element $\gamma_{1} \in {\bf F}_{2^{t_{1}}}$ of order~$m_{1}$;~(2) 
for the odd integer $m_{2}$, 
there exist a finite field ${\bf F}_{2^{t_{2}}}$ with $t_{2} \in [1,m_{2}-1]$ 
and an element $\gamma_{2} \in {\bf F}_{2^{t_{2}}}$~of order $m_{2}$; 
(3) for the odd integer $m=m_{1}m_{2}$, there exist a finite field 
${\bf F}_{2^{t}}$ with $t \in [1,m-1]$ and an element 
$\gamma \in {\bf F}_{2^{t}}$ of order $m$. 
The following lemmas are crucial for our construction. 
\begin{lemma} \label{lemma-subfield}
For the finite fields 
${\bf F}_{2^{t_{1}}}$, ${\bf F}_{2^{t_{2}}}$, and ${\bf F}_{2^{t}}$, 
the following holds$:$ {\rm (1)} ${\bf F}_{2^{t_{1}}}$ is a 
subfield~of~${\bf F}_{2^{t}};$ 
{\rm (2)} ${\bf F}_{2^{t_{2}}}$ is a subfield of 
${\bf F}_{2^{t}};$ {\rm (3)} $t = {\rm lcm}(t_{1},t_{2})$. 
\end{lemma}
{\bf Proof:} For the statement (1), 
it is immediate that ${\bf F}_{2^{t_{1}}}$ 
is a subfield of ${\bf F}_{2^{t}}$ iff $t$ is divisible by~$t_{1}$. 
Note that $t_{1} \in [1,m_{1}-1]$ is a minimum integer 
such that $2^{t_{1}}\equiv 1 \mod{m_{1}}$ and 
$t \in [1,m-1]$ is a minimum integer 
such that $2^{t}\equiv 1 \mod{m}$. Assume that $t$ is not divisible by $t_{1}$, 
i.e., there~exist~$q \geq 1$ and $0 < r < t_{1}$ such that $t=q t_{1}+r$. 
Since $m = m_{1}m_{2}$, we have that $2^{t}\equiv 1 \mod{m_{1}}$.~So~from 
the fact that $2^{t_{1}}\equiv 1 \mod{m_{1}}$, it follows that 
$1 \equiv 2^{t} \equiv 2^{q t_{1}+r} \equiv (2^{t_{1}})^{q} \cdot 2^{r} 
\equiv 2^{r} \mod{m_{1}}.$
%
%
%
%
This contradicts the fact that $t_{1} \in [1,m_{1}-1]$ is a minimum integer 
such that $2^{t_{1}}\equiv 1 \mod{m_{1}}$. Thus $t$ is divisible by $t_{1}$, 
which 
completes the proof of the statement (1). 
The proof of the statement~(2)~is~analogous~to that of the statement (1). 
The statement (3) follows 
from the statements (1) and (2) and the fact that 
$t \in [1,m-1]$ is a minimum integer 
such that $2^{t}\equiv 1 \mod{m}$. \BQED\medskip

For the finite field ${\bf F}_{2^{t_{1}}}$ and 
the element $\gamma \in {\bf F}_{2^{t}}$ given by Lemma \ref{lemma-subfield}, 
the following claims hold: 
\begin{claim} \label{claim-all}
For every 
$h \in {\bf Z}_{m_{1}}^{*}$, $\gamma^{hm_{2}} \in {\bf F}_{2^{t_{1}}}$ is an 
element of order $m_{1}$. 
\end{claim}
{\bf Proof:} Since $2^{t_{1}}\equiv 1 \mod{m_{1}}$, there exists $q \geq 1$ 
such that $2^{t_{1}}-1=qm_{1}$. From the fact~that~$\gamma \in {\bf F}_{2}^{t}$ 
is an element of order $m=m_{1}m_{2}$, 
we have that for every $h \in {\bf Z}_{m_{1}}^{*}$, 
\[
\left(\gamma^{hm_{2}}\right)^{2^{t_{1}}-1} = 
\left(\gamma^{hm_{2}}\right)^{qm_{1}} 
= \left(\gamma^{m_{1}m_{2}}\right)^h=
\left(\gamma^{m}\right)^h=1, 
\]
which implies that $\gamma^{hm_{2}} \in {\bf F}_{2^{t_{1}}}$. It is immediate that 
$(\gamma^{hm_{2}})^{m_{1}}=(\gamma^{m_{1}m_{2}})^{h}=(\gamma^{m})^{h}=1$. 
By~contra\-diction, 
we show that for every $h \in {\bf Z}_{m_{1}}^{*}$, the order of 
$\gamma^{hm_{2}} \in {\bf F}_{2^{t_{1}}}$ is $m_{1}$ 
Assume that there exists~an $h \in {\bf Z}_{m_{1}}^{*}$ such that the order of 
$\gamma^{hm_{2}}$ is $0<\ell<m_{1}$, i.e., 
$(\gamma^{hm_{2}})^{\ell}=\gamma^{h\ell m_{2}}=1$.~Since~the~order~of~$\gamma 
\in {\bf F}_{2^{t}}$ is $m$, 
we have that $h \ell m_{2}$ is divisible~by $m=m_{1}m_{2}$, i.e., 
$h\ell$ is divisible by $m_{1}$. 
From~the~fact~that $h \in {\bf Z}_{m_{1}}^{*}$, 
it follows that $\ell$ is divisible by $m_{1}$, which 
contradicts the assumption that $0 < \ell < m_{1}$. \BQED 
\begin{claim} \label{claim-cover}
In the finite field ${\bf F}_{2^{t_{1}}}$, 
there exist exactly $|{\bf Z}_{m_{1}}^{*}|$ elements of order $m_{1}$. 
\end{claim}
{\bf Proof:} For an element $g \in {\bf F}_{2^{t_{1}}}$ of order $2^{t_{1}}-1$, 
we have that $\alpha = g^{(2^{t_{1}}-1)/m_{1}} \in {\bf F}_{2^{t_{1}}}$ is an 
element~of order $m_{1}$. 
So the $m_{1}$ elements $\alpha^{0},\alpha^{1},\ldots,\alpha^{m_{1}-1}$ are the set of 
all elements that satisfies $x^{m_{1}}=1$.~It~is immediate that 
for each $j \in {\bf Z}_{m}$, the order of $\alpha^{j}$ is $m_{1}/\gcd(j,m_{1})$. 
This~implies~that~in~the~finite~field ${\bf F}_{2^{t_{1}}}$, 
there exist exactly 
$|{\bf Z}_{m_{1}}^{*}|$ elements of order $m_{1}$. \BQED\medskip

In a way similar to the proofs of Claims \ref{claim-all} and \ref{claim-cover}, 
we can also show the following claims~for~the 
finite field ${\bf F}_{2^{t_{2}}}$ and the element 
$\gamma \in {\bf F}_{2^{t}}$ determined by Lemma \ref{lemma-subfield}. 
\begin{claim} \label{claim-all-2}
For every 
$h \in {\bf Z}_{m_{2}}^{*}$, $\gamma^{hm_{1}} \in {\bf F}_{2^{t_{2}}}$ is an 
element of order $m_{2}$. 
\end{claim}
\begin{claim} \label{claim-cover-2}
In the finite field ${\bf F}_{2^{t_{2}}}$, 
there exist $|{\bf Z}_{m_{2}}^{*}|$ elements of order $m_{2}$. 
\end{claim}
From Claims \ref{claim-all}, \ref{claim-cover}, \ref{claim-all-2}, and 
\ref{claim-cover-2}, we can show the following lemma:
\begin{lemma} \label{lemma-element}
For the elements 
$\gamma_{1} \in {\bf F}_{2^{t_{1}}}$, $\gamma_{2} \in {\bf F}_{2^{t_{2}}}$, and 
$\gamma \in {\bf F}_{2^{t}}$, the following holds$:$ 
{\rm (1)} there~exists $h_{1} \in {\bf Z}_{m_{1}}^{*}$ such that 
$\gamma_{1}=\gamma^{h_{1}m_{2}};$ 
{\rm (2)} there exists $h_{2} \in {\bf Z}_{m_{2}}^{*}$ such that 
$\gamma_{2}=\gamma^{h_{2}m_{1}}$. 
\end{lemma}
{\bf Proof:} The statement (1) immediately follows from Claims \ref{claim-all} 
and \ref{claim-cover} and the statement~(2)~immediately follows from 
Claims \ref{claim-all-2} and \ref{claim-cover-2}. \BQED\medskip\\
Let $S_{m_{1}}=\{s_{1}^{1},s_{2}^{1},\ldots,s_{2^{r}-1}^{1}\}$, 
$S_{m_{2}}=\{s_{1}^{2},s_{2}^{2},\ldots,s_{2^{\ell}-1}^{2}\}$, and 
$S_{m}=\{s_{1},s_{2},\ldots,s_{2^{r+\ell}-1}\}$ 
be the canonical sets of the integers $m_{1}$, $m_{2}$, and $m$, respectively, 
and let $s_{0}^{1}=s_{0}^{2}=s_{0}=0$. 
\begin{lemma} \label{lemma-S-composition}
For the sets $S_{m_{1}}$, $S_{m_{2}}$, and $S_{m}$, 
the following holds$:$ For any $s \in S_{m}\cup \{0\}$,~{\rm (1)} $s \in S_{m}$~iff 
there exist $s_{i_{1}}^{1} \in S_{m_{1}} \cup \{0\}$ and $s_{i_{2}}^{2} 
\in S_{m_{2}} \cup \{0\}$ such that $s \equiv 
s_{i_{1}}^{1} \mod{m_{1}}$,~$s \equiv s_{i_{2}}^{2} \mod{m_{2}}$, and 
either $s_{i_{1}}^{1}\neq 0$ or $s_{i_{2}}^{2}\neq 0;$ 
{\rm (2)} $s=0$ iff $s\equiv 0 \mod{m_{1}}$ and $s\equiv 0 \mod{m_{2}}$.
%
%
%
%
%
\end{lemma}
{\bf Proof:} It follows from the definitions of 
$S_{m_{1}}$, $S_{m_{2}}$, and $S_{m}$ and the Chinese Remainder Theorem.~\BQED
%
%
%
%
\subsection{Constructions for Query-Efficient Locally Decodable Codes} 
\label{construction}
%
For the integers $m_{1}$, $m_{2}$, and $m$ and the integer $h>0$ 
given by (\ref{eq-h}), 
let $N_{1}=m_{1}^{h}$, $N_{2}=m_{2}^{h}$,~and~$N=m^{h}$, respectively. The following 
is essential to construct query-efficient locally decodable~codes. 
\begin{thm}[{\tt (Composition Theorem)}] \label{thm-compose}
Let $C_{1}: {\bf F}_{2^{t_{1}}}^{n} \rightarrow {\bf F}_{2^{t_{1}}}^{N_{1}}$ 
be a $k_{1}$-query locally decodable~code that has 
an $S_{m_{1}}$-decoding polynomial 
$P_{1}(x) \in {\bf F}_{2^{t_{1}}}[x] \subseteq {\bf F}_{2^{t}}[x]$ 
with $k_{1}$ monomials and 
$C_{2}: {\bf F}_{2^{t_{2}}}^{n} \rightarrow {\bf F}_{2^{t_{2}}}^{N_{2}}$ 
be a $k_{2}$-query locally decodable code that has an 
$S_{m_{2}}$-decoding polynomial 
$P_{2}(x) \in {\bf F}_{2^{t_{2}}}[x] \subseteq {\bf F}_{2^{t}}[x]$ 
with $k_{2}$ monomials. Then we can construct a 
$k$-query locally decodable code 
$C: {\bf F}_{2^{t}}^{n} \rightarrow {\bf F}_{2^{t}}^{N}$ that 
has an $S_{m}$-decoding polynomial 
$P(x) \in {\bf F}_{2^{t}}[x]$ with $k$ monomials, where $k \leq k_{1}k_{2}$. 
\end{thm}
{\bf Proof:} For $m=m_{1}m_{2}$ and $h$ given by (\ref{eq-h}), we define 
$C: {\bf F}_{2^{t}}^{n} \rightarrow {\bf F}_{2^{t}}^{N}$ as follows: 
For~any~vector~$\vec{x}=(x_{1},x_{2},\ldots, x_{n}) \in {\bf F}_{2^{t}}$, 
let $C(\vec{x})=x_{1}C(\vec{e}_{1})+x_{2}C(\vec{e}_{2})+\cdots + x_{n}C(\vec{e}_{n})$, 
where for each $i \in [1,n]$,~$C(\vec{e}_{i})$ is given by (\ref{eq-encoding}). 
For the integer $h_{1} \in {\bf Z}_{m_{1}}^{*}$ determined by 
Lemma \ref{lemma-element}-(1) and the integer 
$h_{2} \in {\bf Z}_{m_{2}}^{*}$~deter\-mined by Lemma \ref{lemma-element}-(2), 
let $P(x)=P_{1}(x^{h_{1}m_{2}})\cdot P_{2}(x^{h_{2}m_{1}})$. It is obvious that 
$P(x)$ is a polynomial with  $k\leq k_{1}k_{2}$ monomials. 
Let $P(x)=a_{0}+a_{1}x^{b_{1}}+\cdots + a_{k-1}x^{k-1}$. 
For each $i \in [1,n]$, a randomized decoding algorithm $D_{i}$ is defined 
exactly the same as Figure \ref{fig-decoding}. 
For each $i \in [1,n]$, we have that 
\begin{eqnarray*}
D_{i}(C(\vec{x})) & = & 
D_{i}(x_{1} C(\vec{e}_{1}) + x_{2} C(\vec{e}_{2})+ \cdots + 
x_{n} C(\vec{e}_{n}))\\
& = & x_{1} D_{i}(C(\vec{e}_{1})) + x_{2} D_{i}(C(\vec{e}_{2})) + \cdots + 
x_{n} D_{i}(C(\vec{e}_{n})). 
\end{eqnarray*}
Thus it suffices to show that 
$\Pr[D_{i}(C(\vec{e}_{i}))=1]=1$
for each $i \in [1,n]$ and 
$\Pr[D_{i}(C(\vec{e}_{j}))=0]=1$~for each $j \in [1,n]\setminus \{i\}$. 
From (\ref{eq-encoding}), it follows that 
for queries $\vec{v},\vec{v}+b_{1}\vec{u}_{i}, 
\ldots,  \vec{v}+b_{k-1}\vec{u}_{i} \in Z_{m}^{h}$, 
\begin{eqnarray}
D_{i}(C(\vec{e}_{i})) & = & \gamma^{-\pair{\vec{u}_{i},\vec{v}}_{m}} \cdot \left(
a_{0} \gamma^{\pair{\vec{u}_{i},\vec{v}}_{m}} + 
a_{1} \gamma^{\pair{\vec{u}_{i},\vec{v} + b_{1}\vec{u}_{i}}_{m}} + \cdots +
a_{k-1} \gamma^{\pair{\vec{u}_{i},\vec{v}+b_{k-1}\vec{u}_{i}}_{m}}\right)\nonumber\\
& = & 
\gamma^{-\pair{\vec{u}_{i},\vec{v}}_{m}} \cdot \left(
a_{0} \gamma^{\pair{\vec{u}_{i},\vec{v}}_{m}} + 
a_{1} \gamma^{\pair{\vec{u}_{i},\vec{v}}_{m}} 
\gamma^{b_{1}\pair{\vec{u}_{i},\vec{u}_{i}}_{m}} + \cdots +
a_{k-1} \gamma^{\pair{\vec{u}_{i},\vec{v}}_{m}} 
\gamma^{b_{k-1}\pair{\vec{u}_{i},\vec{u}_{i}}_{m}}\right)\nonumber\\
& = & a_{0} + 
a_{1} \gamma^{b_{1}\pair{\vec{u}_{i},\vec{u}_{i}}_{m}} + \cdots +
a_{k-1} \gamma^{b_{k-1}\pair{\vec{u}_{i},\vec{u}_{i}}_{m}}\nonumber\\
& = & P\left(\gamma^{\pair{\vec{u}_{i},\vec{u}_{i}}_{m}}\right) 
= P(1) 
= P_{1}(1)\cdot P_{2}(1) = 1;\nonumber\\
D_{i}(C(\vec{e}_{j})) & = & \gamma^{-\pair{\vec{u}_{i},\vec{v}}_{m}}  \cdot \left(
a_{0} \gamma^{\pair{\vec{u}_{j},\vec{v}}_{m}} + 
a_{1} \gamma^{\pair{\vec{u}_{j},\vec{v} + b_{1}\vec{u}_{i}}_{m}} + \cdots +
a_{k-1} \gamma^{\pair{\vec{u}_{j},\vec{v}+b_{k-1}\vec{u}_{i}}_{m}}\right) \nonumber\\
& = & 
\gamma^{-\pair{\vec{u}_{i},\vec{v}}_{m}} \cdot \left(
a_{0} \gamma^{\pair{\vec{u}_{j},\vec{v}}_{m}} + 
a_{1} \gamma^{\pair{\vec{u}_{j},\vec{v}}_{m}} 
\gamma^{b_{1}\pair{\vec{u}_{i},\vec{u}_{j}}_{m}} + \cdots +
a_{k-1} \gamma^{\pair{\vec{u}_{j},\vec{v}}_{m}} 
\gamma^{b_{k-1}\pair{\vec{u}_{i},\vec{u}_{j}}_{m}}\right) \nonumber\\
& = & 
\gamma^{-\pair{\vec{u}_{i},\vec{v}}_{m}} \cdot 
\gamma^{ \pair{\vec{u}_{j},\vec{v}}_{m}} \cdot \left(
a_{0} + 
a_{1} \gamma^{b_{1}\pair{\vec{u}_{i},\vec{u}_{j}}_{m}} + \cdots +
a_{k-1} \gamma^{b_{k-1}\pair{\vec{u}_{i},\vec{u}_{j}}_{m}}\right).\nonumber\\
& = & \gamma^{-\pair{\vec{u}_{i},\vec{v}}_{m}} \cdot 
\gamma^{ \pair{\vec{u}_{j},\vec{v}}_{m}} \cdot 
P\left(\gamma^{\pair{\vec{u}_{i},\vec{u}_{j}}_{m}}\right)\nonumber\\
& = & \gamma^{-\pair{\vec{u}_{i},\vec{v}}_{m}} \cdot 
\gamma^{ \pair{\vec{u}_{j},\vec{v}}_{m}} \cdot 
P_{1}\left(\gamma^{h_{1}m_{2}\pair{\vec{u}_{i},\vec{u}_{j}}_{m}}\right) \cdot 
  P_{2}\left(\gamma^{h_{2}m_{1}\pair{\vec{u}_{i},\vec{u}_{j}}_{m}}\right)\nonumber\\
& = & \gamma^{-\pair{\vec{u}_{i},\vec{v}}_{m}} \cdot 
\gamma^{ \pair{\vec{u}_{j},\vec{v}}_{m}} \cdot 
P_{1}\left(\gamma_{1}^{\pair{\vec{u}_{i},\vec{u}_{j}}_{m}}\right) \cdot 
  P_{2}\left(\gamma_{2}^{\pair{\vec{u}_{i},\vec{u}_{j}}_{m}}\right), \label{eq-down}
\end{eqnarray}
%
%
%
%
%
where (\ref{eq-down}) follows from Lemma \ref{lemma-element}. 
Since ${\cal U}=\{\vec{u}_{1},\vec{u}_{2},\ldots,\vec{u}_{n}\}$ is a 
family of $S_{m}$-matching vectors,~we have that 
$\pair{\vec{u}_{i},\vec{u}_{j}}_{m} \in S_{m}$. 
Thus from Lemma \ref{lemma-S-composition}, it follows that 
there exist $s_{i_{1}}^{1} \in S_{m_{1}} \cup \{0\}$~and~$s_{i_{2}}^{2} 
\in S_{m_{2}} \cup \{0\}$ such that 
$\pair{\vec{u}_{i},\vec{u}_{j}}_{m} \equiv s_{i_{1}}^{1} \mod{m_{1}}$,  
$\pair{\vec{u}_{i},\vec{u}_{j}}_{m} \equiv s_{i_{2}}^{2} \mod{m_{2}}$, 
and either $s_{i_{1}}^{1}\neq 0$~or~$s_{i_{2}}^{2} \neq 0$. 
Recall that $\gamma_{1} \in {\bf F}_{2^{t_{1}}}$ is an element of order $m_{1}$; 
$\gamma_{2} \in {\bf F}_{2^{t_{2}}}$ is an element of order $m_{2}$; 
$P_{1}(x)$~is~an~$S_{m_{1}}$-decoding polynomial; 
$P_{2}(x)$ is an $S_{m_{2}}$-decoding polynomial. 
Then from (\ref{eq-down}), we have that 
\[
P_{1}\left(\gamma_{1}^{\pair{\vec{u}_{i},\vec{u}_{j}}_{m}}\right) 
= P_{1}\left(\gamma_{1}^{s_{i_{1}}^{1}}\right) = 0 ~\bigvee~
P_{2}\left(\gamma_{2}^{\pair{\vec{u}_{i},\vec{u}_{j}}_{m}}\right) 
= P_{2}\left(\gamma_{1}^{s_{i_{2}}^{2}}\right) = 0. 
\]
Thus it follows that $D_{i}(C(\vec{e_{i}}))=1$ for each $i \in [1,n]$ and 
$D_{i}(C(\vec{e}_{j}))=0$ for each $j \in [1,n]\setminus\{i\}$. \BQED
\begin{cor}[(to Theorem \ref{thm-compose})] \label{cor-LDC-subexp}
For any integer $n>1$ and any integer $r>1$, 
there~exists~a~$k\mbox{-}$que\-ry locally decodable code 
$C: {\bf F}_{2^{t}}^{n} \rightarrow {\bf F}_{2^{t}}^{N}$ such that 
$k \leq 3\cdot 2^{r-2}$ and 
\[
N=\exp\left(\exp\left(O\left(\sqrt[r]{ \log n \cdot 
\left(\log \log n\right)^{r-1}}\right)\right)\right)
=\exp\left(n^{O(\left(\log \log n/\log n\right)^{1-1/r})}\right). 
\]
\end{cor}
{\bf Proof:} Efremenko \cite[Example 3.7]{E} showed that 
for an odd integer $m_{1}=511=7 \cdot 73$,~there~exists a 
3-query locally decodable code 
$C_{1}: {\bf F}_{2^{t_{1}}}^{n} \rightarrow {\bf F}_{2^{t_{1}}}^{N_{1}}$ 
that has an $S_{m_{1}}$-decoding polynomial $P_{1}(x) \in {\bf F}_{2^{t_{1}}}[x]$ 
with 3 monomials. For any integer $r>1$, we take 
$m_{2}=p_{1}^{e_{1}}p_{2}^{e_{2}} \cdots p_{r-2}^{e_{r-2}}$ that is 
a product~of~$r-2$~dis\-tinct odd primes such that $\gcd(m_{1},m_{2})=1$, 
and let $m=m_{1}m_{2}$. Efremenko~\cite[Theorem~3.6]{E}~also~de\-rived that 
for any integer $r > 1$, there exists a $k_{2}$-query locally decodable code 
$C_{1}: {\bf F}_{2^{t_{1}}}^{n} \rightarrow {\bf F}_{2^{t_{1}}}^{N_{1}}$~that 
has an $S_{m_{2}}$-decoding polynomial $P_{2}(x) \in {\bf F}_{2^{t_{2}}}[x]$ 
with $k_{2}$ mono\-mials, where $k_{2}\leq 2^{r}$. 
So~from~The\-orem 
\ref{thm-compose}, we can construct a $k$-query locally decodable code 
$C: {\bf F}_{2^{t}}^{n} \rightarrow {\bf F}_{2^{t}}^{N}$ that has an 
$S_{m}$-decoding polynomial $P(x) \in {\bf F}_{2^{t}}[x]$ with 
$k$ monomials, where $k \leq 3 \cdot 2^{r-2}$. \BQED
%
\section{Concluding Remarks} \label{remarks}
%
In this paper, we have shown the Composition Theorem that constructs a 
$k_{1}k_{2}$-query locally decoda\-ble code by composing 
a $k_{1}$-query locally decodable code and a $k_{2}$-query 
locally decodable code (see Theorem \ref{thm-compose}) 
and in Corollary 
\ref{cor-LDC-subexp}, we have also shown that 
for any integer $r>1$, 
there exists~a~$k$-query locally decodable code 
$C: {\bf F}_{2^{t}}^{n} \rightarrow {\bf F}_{2^{t}}^{N}$ such that 
$k \leq 3\cdot 2^{r-2}$ and 
\[
N=\exp\left(\exp\left(O\left(\sqrt[r]{ \log n \cdot 
\left(\log \log n\right)^{r-1}}\right)\right)\right)
=\exp\left(n^{O(\left(\log \log n/\log n\right)^{1-1/r})}\right). 
\]
For perfectly smooth decoders, we can immediately 
modify Theorem \ref{thm-compose} as follows: 
\begin{thm} \label{thm-compose-smooth-decoder}
Let $C_{1}: {\bf F}_{2^{t_{1}}}^{n} \rightarrow {\bf F}_{2^{t_{1}}}^{N_{1}}$ 
be a $k_{1}$-query locally decodable code with a perfectly smooth decoder 
${\cal D}_{1}$ that has an $S_{m_{1}}$-decoding polynomial $P_{1}(x) \in 
{\bf F}_{2^{t_{1}}}[x] \subseteq {\bf F}_{2^{t}}[x]$ with 
$k_{1}$ monomials and 
$C_{2}: {\bf F}_{2^{t_{2}}}^{n} \rightarrow {\bf F}_{2^{t_{1}}}^{N_{2}}$ 
be a $k_{2}$-query locally decodable code with 
a perfectly smooth decoder ${\cal D}_{2}$ that has 
an $S_{m_{2}}$-decoding polynomial 
$P_{2}(x) \in {\bf F}_{2^{t_{2}}}[x] \subseteq {\bf F}_{2^{t}}[x]$ 
with $k_{2}$ monomials. 
Then~we~can~construct~a $k$-query locally decodable code 
$C: {\bf F}_{2^{t}}^{n} \rightarrow {\bf F}_{2^{t}}^{N}$ 
with~a~perfect\-ly smooth decoder ${\cal D}$ 
that~has~an~$S_{m}$-decoding polynomial 
$P(x) \in {\bf F}_{2^{t}}[x]$~with~$k$ monomials, where $k \leq k_{1}k_{2}$. 
\end{thm}
From Theorem \ref{thm-compose-smooth-decoder} and the 
transformation \cite{T} from a $k$-query locally decodable codes~with~a~perfect\-ly 
smooth decoder to $k$-server private information retrieval, 
we can show the following theorem:
\begin{thm} \label{thm-pir}
For any integer $n>1$ and any integer $r>1$, 
there exists a $k$-server private~informa\-tion retrieval 
such that $k \leq 3\cdot 2^{r-2}$ and 
$C_{k}(n) = n^{O\left(\left(\log \log n/\log n\right)^{(r-1)/r}\right)}$. 
\end{thm}

At present, we know only a 3-query locally decodable code 
${\bf F}_{2^{9}}^{n} \rightarrow {\bf F}_{2^{9}}^{N}$ such that 
\[
N=\exp\left(\exp\left(O\left(\sqrt{ \log n \cdot 
\log \log n }\right)\right)\right)
=\exp\left(n^{O(\left(\log \log n/\log n\right)^{1/2})}\right), 
\]
for an add integer $m=511=2^{9}=7 \cdot 73$ \cite[Example 3.7]{E}. 
Let ${\cal M}_{r}$ be a set of 
integers,~each~of~which is a product of $r>1$ distinct odd primes. 
From the Composition Theorem 
(see Theorem \ref{thm-compose}),~it~follows~that if there exist 
$m_{1},m_{2},\dots,m_{\ell} \in {\cal M}_{2}$ such that $\gcd(m_{i},m_{j})=1$ 
for each $1 \leq i < j \leq \ell$~and each $m_{i} \in {\cal M}_{2}$ generates 
a 3-query locally decodable code 
$C_{i}:{\bf F}_{2^{t_{i}}}^{n}\rightarrow {\bf F}_{2^{t_{i}}}^{N}$ that has an 
$S_{m_{i}}$-decoding polyno\-mial $P_{i}(x) \in {\bf F}_{2^{t_{i}}}[x]$ 
with less than 4 monomials, where 
\[
N=\exp\left(\exp\left(O\left(\sqrt{ \log n \cdot 
\log \log n }\right)\right)\right)
=\exp\left(n^{O(\left(\log \log n/\log n\right)^{1/2})}\right), 
\]
then for the integer $m=m_{1}m_{2}\cdots m_{\ell}$, 
we can construct a $k$-query locally decodable code 
$C:{\bf F}_{2^{t}}^{n}\rightarrow {\bf F}_{2^{t}}^{N}$ 
that has an $S_{m}$-decoding polynomial $P(x) \in {\bf F}_{2^{t}}[x]$ 
with $k$ monomials, where $k \leq 3^{\ell}$ and 
\[
N=\exp\left(\exp\left(O\left(\sqrt[2 \ell]{ \log n \cdot 
\log \log n }\right)\right)\right)
=\exp\left(n^{O(\left(\log \log n/\log n\right)^{1-1/2\ell})}\right), 
\]
however, we do not know such integers $m_{1},m_{2},\ldots,m_{\ell} \in {\cal M}_{2}$ 
exist other than $m=511 \in {\cal M}_{2}$.~Thus the following problems are 
both of theoretical interest and of practical importance. 
\begin{namelist}{~~~(1)}
\item[(1)] Find integers $m \in {\cal M}_{2}\setminus \{511\}$ that 
generate a 3-query locally decodable code 
$C:{\bf F}_{2^{t}}^{n}\rightarrow {\bf F}_{2^{t}}^{N}$, i.e., 
the code $C$ has an 
$S_{m}$-decoding polynomial $P(x) \in {\bf F}_{2^{t}}[x]$ 
with less than 4 monomials.
\item[(2)] For any integer $r>2$, find an integer $m \in {\cal M}_{r}$ 
that generate a $k$-query locally decodable code 
$C:{\bf F}_{2^{t}}^{n}\rightarrow {\bf F}_{2^{t}}^{N}$ 
that has an 
$S_{m}$-decoding polynomial $P(x) \in {\bf F}_{2^{t}}[x]$ 
with $k < 3 \cdot 2^{r-2}$ monomials.
\end{namelist}
%
\subsection*{Acknowledgments}
%
The authors would like to thank Osamu Watanabe for his insightful discussions and 
valuable comments for the earlier version of the paper. 
%

%
\appendix
%
\section{Proof of Lemma \ref{lemma-smooth}} \label{appendix-proof-smooth}
%
For each $i \in [1,n]$, 
it is obvious that each of queries $\vec{v},\vec{v}+b_{1}\vec{u}_{i}, 
\ldots,  \vec{v}+b_{k-1}\vec{u}_{i} \in Z_{m}^{h}$~is~uniformly~dis\-tributed 
over $[1,N]$. 
So for any vector $\vec{x}=(x_{1},x_{2},\ldots,x_{n}) \in {\bf F}_{2^{t}}^{n}$, 
we show that $\Pr[D_{i}(C(\vec{x}))=x_{i}]=1$ for each $i \in [1,n]$. 
Since $C(\vec{x}) = 
x_{1} C(\vec{e}_{1})+ x_{2} C(\vec{e}_{2})+ \cdots + 
x_{n} C(\vec{e}_{n})$, 
we have that~for~each~$i \in [1,n]$, 
\begin{eqnarray*}
D_{i}(C(\vec{x})) & = & 
D_{i}(x_{1} C(\vec{e}_{1}) + x_{2} C(\vec{e}_{2})+ \cdots + 
x_{n} C(\vec{e}_{n}))\\
& = & x_{1} D_{i}(C(\vec{e}_{1})) + x_{2} D_{i}(C(\vec{e}_{2})) + \cdots + 
x_{n} D_{i}(C(\vec{e}_{n})). 
\end{eqnarray*}
Thus it suffices to show that 
$\Pr[D_{i}(C(\vec{e}_{i}))=1]=1$
for each $i \in [1,n]$ and 
$\Pr[D_{i}(C(\vec{e}_{j}))=0]=1$~for each $j \in [1,n]\setminus \{i\}$. 
From (\ref{eq-encoding}), it follows that 
for queries $\vec{v},\vec{v}+b_{1}\vec{u}_{i}, 
\ldots,  \vec{v}+b_{k-1}\vec{u}_{i} \in Z_{m}^{h}$, 
\begin{eqnarray}
D_{i}(C(\vec{e}_{i})) & = & \gamma^{-\pair{\vec{u}_{i},\vec{v}}_{m}} \cdot \left(
a_{0} \gamma^{\pair{\vec{u}_{i},\vec{v}}_{m}} + 
a_{1} \gamma^{\pair{\vec{u}_{i},\vec{v} + b_{1}\vec{u}_{i}}_{m}} + \cdots +
a_{k-1} \gamma^{\pair{\vec{u}_{i},\vec{v}+b_{k-1}\vec{u}_{i}}_{m}}\right)\nonumber\\
& = & 
\gamma^{-\pair{\vec{u}_{i},\vec{v}}_{m}} \cdot \left(
a_{0} \gamma^{\pair{\vec{u}_{i},\vec{v}}_{m}} + 
a_{1} \gamma^{\pair{\vec{u}_{i},\vec{v}}_{m}} 
\gamma^{b_{1}\pair{\vec{u}_{i},\vec{u}_{i}}_{m}} + \cdots +
a_{k-1} \gamma^{\pair{\vec{u}_{i},\vec{v}}_{m}} 
\gamma^{b_{k-1}\pair{\vec{u}_{i},\vec{u}_{i}}_{m}}\right)\nonumber\\
& = & a_{0} + 
a_{1} \gamma^{b_{1}\pair{\vec{u}_{i},\vec{u}_{i}}_{m}} + \cdots +
a_{k-1} \gamma^{b_{k-1}\pair{\vec{u}_{i},\vec{u}_{i}}_{m}}\nonumber\\
& = & P\left(\gamma^{\pair{\vec{u}_{i},\vec{u}_{i}}_{m}}\right);\label{eq-app-1}\\
D_{i}(C(\vec{e}_{j})) & = & \gamma^{-\pair{\vec{u}_{i},\vec{v}}_{m}} \cdot \left(
a_{0} \gamma^{\pair{\vec{u}_{j},\vec{v}}_{m}} + 
a_{1} \gamma^{\pair{\vec{u}_{j},\vec{v} + b_{1}\vec{u}_{i}}_{m}} + \cdots +
a_{k-1} \gamma^{\pair{\vec{u}_{j},\vec{v}+b_{k-1}\vec{u}_{i}}_{m}}\right) \nonumber\\
& = & 
\gamma^{-\pair{\vec{u}_{i},\vec{v}}_{m}} \cdot \left(
a_{0} \gamma^{\pair{\vec{u}_{j},\vec{v}}_{m}} + 
a_{1} \gamma^{\pair{\vec{u}_{j},\vec{v}}_{m}} 
\gamma^{b_{1}\pair{\vec{u}_{i},\vec{u}_{j}}_{m}} + \cdots +
a_{k-1} \gamma^{\pair{\vec{u}_{j},\vec{v}}_{m}} 
\gamma^{b_{k-1}\pair{\vec{u}_{i},\vec{u}_{j}}_{m}}\right) \nonumber\\
& = & 
\gamma^{-\pair{\vec{u}_{i},\vec{v}}_{m}} \cdot
\gamma^{ \pair{\vec{u}_{j},\vec{v}}_{m}} \cdot \left(
a_{0} + 
a_{1} \gamma^{b_{1}\pair{\vec{u}_{i},\vec{u}_{j}}_{m}} + \cdots +
a_{k-1} \gamma^{b_{k-1}\pair{\vec{u}_{i},\vec{u}_{j}}_{m}}\right).\nonumber\\
& = & \gamma^{-\pair{\vec{u}_{i},\vec{v}}_{m}} \cdot
\gamma^{ \pair{\vec{u}_{j},\vec{v}}_{m}} \cdot
P\left(\gamma^{\pair{\vec{u}_{i},\vec{u}_{j}}_{m}}\right).\label{eq-app-2}
%
\end{eqnarray}
Since ${\cal U}=\{\vec{u}_{1}, \vec{u}_{2}, \ldots, \vec{u}_{n}\}$  is a 
family of $S_{m}$-matching vectors, 
we have that~$\pair{\vec{u}_{i},\vec{u}_{i}}_{m}=0$~for~each~$i \in [1,n]$ and 
$\pair{\vec{u}_{i},\vec{u}_{j}}_{m} = s_{ij} \in S_{m} \subseteq Z_{m}\setminus \{0\}$ 
for each $i,j \in [1,n]$ 
such that $i \neq j$,~and~from~the~definition of $S_{m}$-decoding polynomial 
$P(x)=a_{0}+a_{1}x^{b_{1}}+\cdots + a_{k-1}x^{b_{k-1}}$, 
we have that $P(\gamma^{\pair{\vec{u}_{i}, \vec{u}_{i}}_{m}})=P(1)=1$ for 
each $i \in [1,n]$ and 
$P(\gamma^{\pair{\vec{u}_{i}, \vec{u}_{j}}_{m}})=P(\gamma^{s_{ij}})=0$ for each 
$i,j \in [1,n]$ such that $i \neq j$. 

Thus~it~follows from (\ref{eq-app-1}) that 
$D_{i}(C(\vec{e}_{i})) = 
P(\gamma^{\pair{\vec{u}_{i},\vec{u}_{i}}_{m}})= P(1)=1$ 
for each $i \in [1,n]$,~and~it~follows from (\ref{eq-app-2}) that 
$D_{i}(C(\vec{e}_{j})) = 
\gamma^{-\pair{\vec{u}_{i},\vec{v}}_{m}} \cdot 
\gamma^{ \pair{\vec{u}_{j},\vec{v}}_{m}} \cdot P(\gamma^{s_{ij}})=0$
for each $i,j \in [1,n]$~such~that~$i \neq j$. 

\begin{thebibliography}{99}
%
\bibitem{A} A. Ambainis. Upper Bound on the Communication Complexity of 
Private Information Retrieval. In {\it Proc. of the 24th International 
Colloquium on Automata, Languages, and Programming\/}, Lecture Notes in 
Computer Science 1256, pp.401-407 (1997).
%
\bibitem{BFLS} L. Babai, L. Fortnow, L. Levin, and M. Szegedy. 
Checking Computation in Polylogarithmic Time. 
In {\it Proc. of 
the 23rd Annual ACM Symposium on Theory of Computing\/}, pp.21-31 (1991). 
%
\bibitem{BFG} A. Beimel, L. Fortnow, and W. Gasarch. A Tight Lower Bound 
for Restricted PIR Protocols. {\it Comoutat. Comlex.\/}, 15, pp.82-91 (2006). 
%
\bibitem{BIK} A. Beimel, Y. Ishai, and E. Kushilevitz. General Constructions 
for Information-Theoretic Private Information Retrieval. 
{\it J. of Computer and System Sciences\/}, 71(2), pp.213-247 (2005). 
%
\bibitem{BIKR} A. Beimel, Y. Ishai, E. Kushilevitz, and F. Raymond. 
Breaking the $O(n^{\frac{1}{2k-1}})$~Barrier~for~Infor\-mation-Theoretic 
Private Information Retrieval. In {\it Proc. of the 43rd IEEE Annual Symposium 
on Foundations of Computer Science\/}, pp.261-270 (2002). 
%
\bibitem{CGKS} B. Chor, O. Goldreich, E. Kushilevitz, and M. Sudan. 
Private Information Retrieval. In {\it Proc. of the 36th IEEE Annual Symposium on 
Foundations of Computer Science\/}, pp.41-51 (1995). 
%
\bibitem{E} K. Efremenko. 3-Query Locally Decodable Codes of Subexponential Length. 
{\it Electronic Colloquium on Computational Complexity\/}, Report No.69 (2008). 
%
\bibitem{Ga} W. Gasarch. A Survey on Private Information Retrieval. 
{\it Bull. EATCS\/} 82, pp.72-107 (2004). 
%
\bibitem{G} V. Grolmusz. Superpolynomial Size Set-Systems with Restricted 
Intersections Mod 6 and Explicit Ramsey Graphs. {\it Combinatorica\/}, 
20(1), pp.71-85 (2000).
%
\bibitem{Go} O. Goldreich. Short Locally Testable Codes and Proofs. 
{\it Electronic Colloquium on Computational Complexity\/}, Report No.14 (2005). 
%
\bibitem{GKST} O. Goldreich, H. Karloff, L.J. Schulman, and L. Trevisan. 
Lower Bounds for Linear Locally Decodable Codes and Private Information 
Retrieval. In {\it Proc. of the 17th IEEE Annual Conference on Computational 
Complexity\/}, pp.175-183 (2002). 
%
\bibitem{IK} Y. Ishai and E. Kushilevitz. Improved Upper Bounds on 
Information-Theoretic Private Information Retrieval. In {\it Proc. of 
the 31st Annual ACM Symposium on Theory of Computing\/}, pp.79-88 (1999). 
%
\bibitem{I1} T. Itoh. Efficient Private Information Retrieval. 
{\it IEICE Trans. Fund. Electron. Commun. Comput. Sci.\/}, 
E82-A(1), pp.11-20 (1999).
%
\bibitem{I2} T. Itoh. On Lower Bound for the Communication Complexity of 
Private Information Retrieval. 
{\it IEICE Trans. Fund. Electron. Commun. Comput. Sci.\/}, 
E84-A(1), pp.157-164 (2001).
%
\bibitem{KW} I. Kerenidis and R. de Wolf. Exponential Lower Bound for 
2-Query Locally Decodable Code via a Quantum Argument. In 
{\it Proc. of the 35th Annual ACM Symposium on Theory of Computing\/}, 
pp.106-115 (2003).
%
\bibitem{KT} J. Katz and L. Trevisan. On the Efficiency of Locally Decoding 
Procedures for Error-Correcting Codes. In {\it Proc. of 
the 32nd Annual ACM Symposium on Theory of Computing\/}, 
pp.80-86 (2000).
%
%
\bibitem{M} E. Mann. Private Access to Distributed Information. Master's 
Thesis, Technion -- Israel Institute of Technology, Haifa, Israel (1998). 
%
\bibitem{PS} A. Polisgchuk and D. Spielman. Nearly-Linear Size Holographic Proofs. 
In {\it Proc. of the 26th Annual ACM Symposium on Theory of Computing\/}, 
pp.194-203 (1994).
%
\bibitem{RY} A. Razborov and S. Yekhanin. An $\Omega(n^{1/3})$ Lower Bounds 
for Bilinear Group Based Private Information Retrieval. 
In {\it Proc. of the 47th Annual IEEE Symposium 
on Foundations of Computer Science\/}, pp.739-748 (2006). 
{\it \/}
%
%
\bibitem{S} M. Sudan. Efficient Checking of Polynomials and Proofs and 
the Hardness of Approximation Problems. Ph.D Thesis, University of 
California at Berkeley (1992). 
%
\bibitem{T} R. Trevisan. Some Applications of Coding Theory in Computational 
Complexity. {\it Quad. Matemat.\/} 13, pp.347-424 (2004). 
{\it Electronic Colloquium on Computational Complexity\/}, Report No.43 (2004). 
%
\bibitem{WW} S. Wehner and R. de Wolf. Improved Lower Bound for Locally 
Decodable Codes and Private Information Retrieval. In {\it Proc. of 
the 32nd International Colloquium on Automata, Languages, and Programming\/}, 
Lecture Notes in Computer Science 3580, pp.1424-1436 (1997).
%
\bibitem{W} D. Woodruff. New Lower Bounds for General Locally Decodable Codes. 
{\it Electronic Colloquium on Computational Complexity\/}, Report No.6 (2007). 
%
\bibitem{WY} D. Woodruff and S. Yekhanin. A Geometric Approach to Information 
Theoretic Private Information Retrieval. {\it SIAM J. Comput.\/}, 
37(4), pp.1046-1056 (2007). 
%
\bibitem{Y1} S. Yekhanin. Towards 3-Query Locally Decodable Codes of 
Subexponential Length. In {\it Proc. of the 39th Annual ACM Symposium on 
Theory of Computing\/}, pp.266-274 (2007).
%
\bibitem{Y2} S. Yekhanin. Towards 3-Query Locally Decodable Codes of 
Subexponential Length. {\it J. of the ACM\/}, 55(1), pp.1-16 (2008). 
%
\end{thebibliography}
\end{document}